\documentclass[twocolumn,aps,prl,amssymb,amstext,showpacs]{revtex4}
\usepackage{graphicx,float,amssymb,amstext,amsmath,dsfont}
\usepackage{dcolumn}
\usepackage{bm}

\newcommand{\vm}{{\mathbf{m}}}
\newcommand{\vn}{{\mathbf{n}}}
\newcommand{\vJ}{{\mathbf{J}}}
\newcommand{\vmu}{{\boldsymbol{\mu}}}
\newcommand{\valpha}{{\boldsymbol{\alpha}}}
\newcommand{\vomega}{{\boldsymbol{\Omega}}}
\newcommand{\vphi}{\boldsymbol{\varphi}}
\newcommand{\ve}{\mathbf{e}}
\newcommand{\va}{\mathbf{a}}

\newcommand{\sign}{\operatorname{sign}}

\newcommand{\ud}{\mathrm{d}}
\newcommand{\ui}{\mathrm{i}}
\newcommand{\ue}{\mathrm{e}}

\newcommand{\R}{\mathds{R}}
\newcommand{\Z}{\mathds{Z}}
\newcommand{\N}{\mathds{N}}

\newcommand{\pa}{\partial}

\providecommand{\abs}[1]{\lvert#1\rvert}

\begin{document}

\title{A periodic orbit formula for quantum reactions through transition states}

\author{Roman Schubert$^1$} 

\author{Holger Waalkens$^{1,2}$} 

\author{Arseni Goussev$^1$} 

\author{Stephen Wiggins$^{1}$} 
\affiliation{$^1$School of Mathematics, University of Bristol, University Walk, Bristol BS8 1TW, UK\\
$^2$Johann Bernoulli Institute for Mathematics and Computer Sciences, University of Groningen, PO Box 407, 
9700 AK Groningen, the Netherlands}

\date{\today}

\begin{abstract}

  Transition State Theory forms the basis of computing reaction rates
  in chemical and other systems.  Recently it has been shown how
  transition state theory can rigorously be realized in phase space
  using an explicit algorithm. The quantization has been demonstrated
  to lead to an efficient procedure to compute cumulative reaction
  probabilities and the associated Gamov-Siegert resonances. In this
  letter these results are used to express the cumulative reaction
  probability as an absolutely convergent sum over periodic orbits
  contained in the transition state. 

\end{abstract}


\maketitle


\emph{Introduction.---}
Transition State Theory, developed by Eyring, Polanyi and Wigner in
the 1930's, is the most fundamental and widely used method to compute
reaction rates. During a reaction a molecular system is envisaged to
pass through a `transition state' or `activated complex', a kind of
unstable supermolecule poised between reactants and products
\cite{Pechukas81}. The main idea of transition state theory is to
place a dividing surface in the transition state region and compute
the classical reaction rate from the directional flux through the
dividing surface. In order not to overestimate the reaction rate the
dividing surface needs to have the crucial property that it is crossed
exactly once by all reactive trajectories (trajectories passing from
reactants to products or vice versa) and not crossed at all by all
other (non-reactive) trajectories. In the 1970's Pechukas, Pollak and
others showed how to rigorously construct such a dividing surface from
a periodic orbit giving the so-called \emph{periodic orbit dividing
  surface} (PODS) \cite{PechukasMcLafferty73}. The generalization to
more degrees of freedoms has posed a major problem, and was solved
only recently using ideas from dynamical systems theory (see
\cite{WWJU01}). This shows that the transition state at energy $E$ is
formed by a \emph{normally hyperbolic invariant manifold} (NHIM) (see
\cite{Wiggins94}), which in this case is an invariant sphere of
dimension $2d-3$, where $d$ is the number of degrees of freedoms, and
normal hyperbolicity means that the contraction and expansion rates
associated with the directions normal to the sphere dominate those of
the directions tangential to the sphere. For $d=2$, this simply is the
unstable periodic orbit of the PODS \cite{WaalkensWiggins04}. In fact,
the NHIM spans another sphere which is of dimension $2d-2$ and hence
has one dimension less than the energy surface and can be taken as a
dividing surface. The NHIM forms the equator of this sphere and
divides it into one hemisphere crossed exactly once by all forward
reactive trajectories and one hemisphere crossed exactly once by all
backward reactive trajectories.  The NHIM itself is invariant and can
be viewed as the energy surface of an invariant subsystem (the
`transition state' or `activated complex') with one degree of freedom
less than the full system (i.e. with the reaction coordinate being
frozen at a particular value).  All these phase space structures can
be explicitly constructed from a \emph{normal form} which at the same
time gives a simple expression for the flux through the dividing
surface.
In \cite{SchubertWaalkensWiggins06} the quantization of this normal
form has been used to develop a quantum version of transition state
theory. This \emph{quantum normal form} has been demonstrated to give
an efficient method to compute cumulative reaction probabilities (the
quantum analogue of the classical flux) and Gamov-Siegert resonances
associated with the activated complex
\cite{SchubertWaalkensWiggins06,GoussevSchubertWaalkensWiggins2009}. In
this letter we use these results to show that the cumulative reaction
probability can be expressed as a sum over periodic orbits contained
in the activated complex.

\emph{The normal form representation of the activated complex and the computation of reaction rates.---}
We consider a molecular system with $d=1+f$ degrees of freedom
which has a saddle-center-\ldots-center equilibrium point (`saddle'
for short), i.e. the matrix associated with the linearized Hamilton's
equations has one pair of real eigenvalues $\pm \lambda$, and $f$
pairs of purely imaginary eigenvalues $\pm \ui \omega_k$,
$k=1,\ldots,f$. We will restrict ourselves to the generic case of
linear frequencies $\omega_k$ fulfilling no resonance condition
$m_1\omega_1+\ldots+m_f\omega_f=0$ for any vector of integers $\vm =
(m_1,\ldots,m_f)\ne 0$. Such saddles are characteristic for reaction
type dynamics as for energies near the energy of the saddle, they
induce a bottleneck type structure of the energy surface near the
saddle through which the system has to pass in order to react.
 
\begin{figure}
\begin{center}
\includegraphics[angle=0,width=5cm]{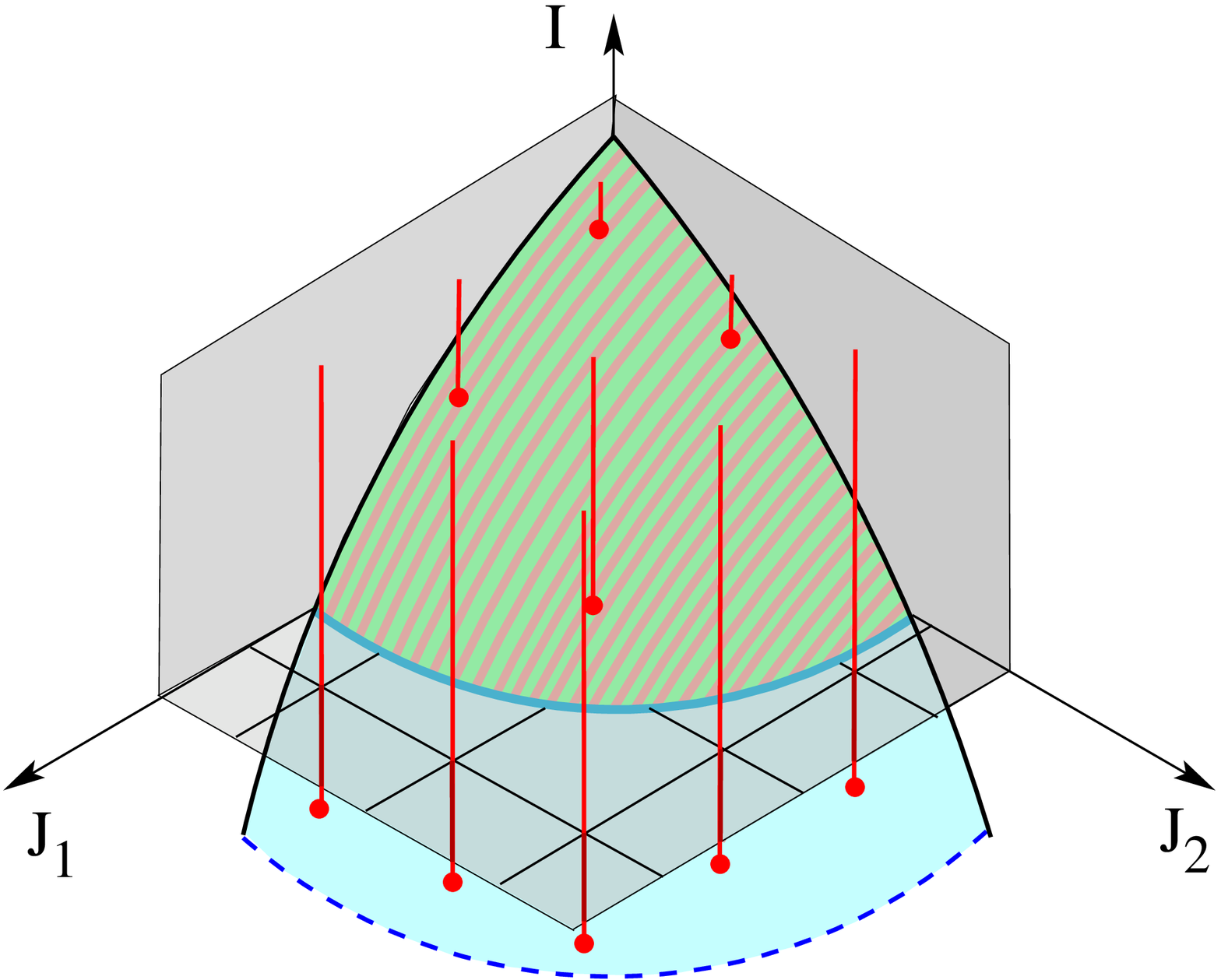}
\includegraphics[angle=0,width=3cm]{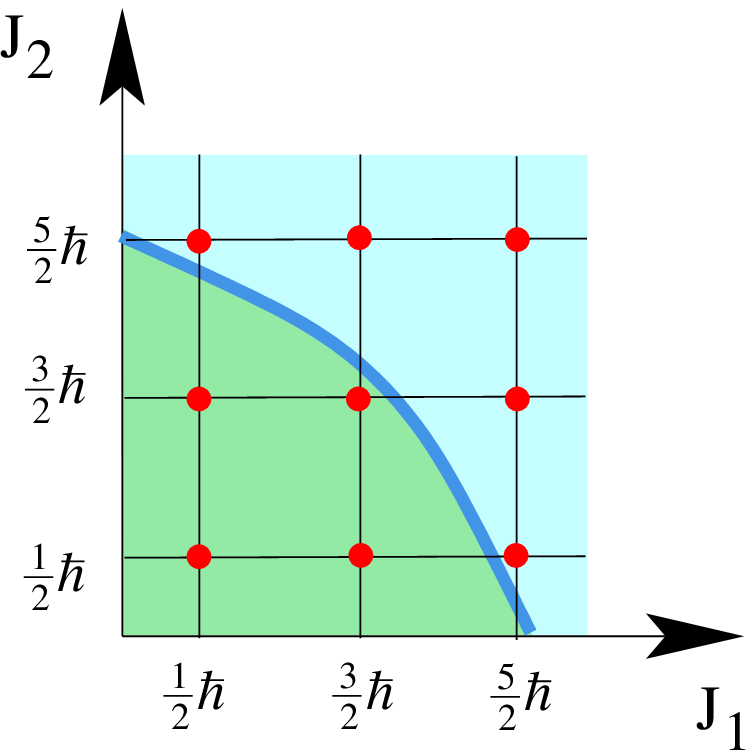}
\end{center}
\caption{\label{fig:momentummap}
For $d=3$ degrees of freedom, the left panel shows an
energy surface $H_0(I,J_2,J_3)=E$ for an energy above the saddle energy. The red lines mark the Bohr-Sommerfeld quantized actions $\vJ$. 
The right panel  shows the energy surface of the activated complex $\Sigma_E$ defined in \eqref{eq:def_Sigma_E} marked as the blue line in the left panel. The enclosed area is proportional to the classical flux, and equivalently, to the mean number of states of the activated complex.
}
\end{figure}

Normal form theory shows that in the neighborhood of the saddle there
is a canonical transformation such that the transformed Hamiltonian is
of the form $H_0(I,J_1,\ldots,J_f)$, where $I=(p_0^2-q_0^2)/2$ is an
integral associated with the reaction coordinate, and the
$J_k=(p_k^2+q_k^2)/2$, $k=1,\ldots,f$, are action integrals associated
with the bath modes. The activated complex is the invariant subsystem
given by $p_0=q_0=0$.  Its motions are described by the reduced
Hamiltonian $H_0(0,J_1,\ldots,J_f)$, and thus is integrable, i.e. in
action angle variables $(\vJ,\vphi)$ the equations of motion are
$\dot{\vJ}=0$ and ${\dot{\vphi}}=\nabla_\vJ H_0(0,\vJ)$ with solutions
$\vJ(t)=const$ and
\begin{equation}\label{eq:orbits}
\vphi(t)=\vphi_0+t\,\vomega(\vJ) \text{ mod } 2\pi \,,\,\text{where}\,\, \vomega(\vJ):=
\nabla_{\vJ}H_0(0,\vJ)\,\, .
\end{equation}
The motion is thus quasiperiodic. It takes place on invariant $f$ dimensional Liouville-Arnold tori \cite{Arnold78} which foliate 
the phase space of the activated complex.
The motion becomes  periodic for the $\vJ$  for which 
$\vomega(\vJ)= a\vm$, where $\vm\in \Z^f$ and  $a\in \R$. We call the torus corresponding to this $\vJ$ a \emph{resonant torus}. 
Fixing the energy $E$ the energy surface of the activated complex,
\begin{equation}\label{eq:def_Sigma_E}
\Sigma_E=\{ \vJ \in \R^f_+ \,: \, H_0(0,J_1,\ldots,J_f) = E \}\,,
\end{equation}
is the action space projection of the NHIM mentioned in the
introduction. The volume it encloses in the space of the actions $\vJ$
is proportional to the directional flux through the dividing surface
(see Fig.~\ref{fig:momentummap}).

A quantum normal form procedure based on the Weyl symbol calculus
\cite{SchubertWaalkensWiggins06} shows that in the quantum mechanical
case a unitary transformation can be found which transforms the
Hamilton operator to the form
$\hat{H}=H(\hat{I},\hat{J}_1,\ldots,\hat{J}_n)$ which is a polynomial
function of the operators $\hat{I}=(-\hbar^2\partial_{q_1}^2
-q_1^2)/2$ and $\hat{J}_k=(-\hbar^2\partial_{q_k}^2 +q_k^2)/2$
associated with the classical integrals.  The polynomial defining the
quantum normal form operator has the $\hbar$ expansion
$H(I,\vJ)=H_0(I,\vJ)+\hbar H_1(I,\vJ)+\ldots$, where $H_k(I, \vJ)$ are
independent of $\hbar$, and $H_0(I,\vJ)$ coincides with the classical
normal form Hamiltonian.

The cumulative reaction probability at energy $E$ is then given by
\begin{equation} \label{eq:def_cum_reac_prob}
N(E) = \sum_{{\bf n}\in\N_0^f} \frac{1}{1+\ue^{-2\pi I_{\bf n}/\hbar}}\,,
\end{equation}
where $I_\vn=I_\vn(E)$ is implicitly defined by 
\begin{equation}\label{eq:energyequation}
H(I_{\bf n},J_1,\ldots,J_f)=E\,,
\end{equation}
and $\vn =(n_1,\ldots,n_f)\in \N_0^f$ is the vector of quantum numbers for the Bohr-Sommerfeld quantized actions,
\begin{equation}
J_k=\hbar(n_k+\frac{\alpha_k}{2})\,,\quad k=1,\ldots,f\,.
\end{equation}
Here the $\alpha_k=2$ are Maslov indices which for later reference we group in the vector $\valpha=(\alpha_1,\ldots,\alpha_f)$
(see \cite{Miller77} for an earlier reference and \cite{SchubertWaalkensWiggins06} where this result is derived in a systematic semiclassical expansion in $\hbar$).
In the following we derive a  formula which expresses $N(E)$ in terms of a sum over periodic orbits.


\emph{A periodic orbit formula for the cumulative reaction
  probability.---} To derive our periodic orbit formula it is
convenient to consider the energy derivative of the cumulative
reaction probability \eqref{eq:def_cum_reac_prob},
\begin{equation} \label{eq:def_n}
n(E) := \frac{\ud N(E)}{\ud E} 
=  \sum_{{\bf n}\in\N_0^f}  \frac{2\pi}{\hbar}  \frac{\ud I_n}{\ud E}  \frac{1}{4\cosh^2(\pi I_\vn/\hbar)} \,.
\end{equation} 
Using  \eqref{eq:energyequation} the factor $\ud I_n/\ud E$ can be written as
\begin{equation}
\frac{\ud I_\vn}{\ud E} =  \left. \frac{\partial H}{\partial I}  \right|_{I=I_\vn,J_k=\hbar(n_k+\frac12)}  ^{-1}\,.
\end{equation}
We can obtain a periodic orbit formula for $n(E)$ following a
computation similar to the derivation of the Berry-Tabor trace formula
for the density of states of classically integrable systems
\cite{BerryTabor76}. Following \cite{BerryTabor76} we use the Poisson
summation formula to rewrite \eqref{eq:def_n} as
\begin{equation}
\begin{split}
\label{eq:def_n_m}   
   n(E) 
     =& \sum_{{\vm}\in\Z^f} n_{\bf m}(E):=  \sum_{{\bf m}\in\Z^f} \frac{2\pi}{\hbar^{f+1}}   
     \ue^{-\ui \pi \valpha \vm/2} \times \\ 
 &  \int \ud^f J    
    \left. \frac{\partial H}{\partial I} \right|_{I=I(E,\vJ)}^{-1} 
     \frac{1}{4\cosh^2 (\pi I(E,\vJ)/\hbar)}      
     \ue^{2\pi\ui {\bf m}\cdot {\bf J} / \hbar} \,,
     \end{split}
\end{equation}
where $I(E,\vJ)$ is determined by 
\begin{equation}\label{eq:energyequation2}
H(I(E,\vJ),J_1,\ldots,J_f)=E\,.
\end{equation} 
Note that the $\hbar$ expansion of the quantum normal form Hamiltonian
implies, via \eqref{eq:energyequation2}, an $\hbar$ expansion of
$I(E,\vJ)$, i.e.  $I(E,\vJ)=I_0(E,\vJ)+\hbar I_1(E,\vJ)+\ldots$.
In the following we separately discuss the term $n_0$ which we refer to as the Thomas-Fermi term \cite{BerryTabor76}, and the remaining sum over 
$\vm\ne 0$ which we refer to as the oscillatory term $n_{\text{osc}}(E)$.

\emph{The Thomas-Fermi term.---}
For ${\bf m}=0$, we get 
\begin{equation}
n_0(E) = \frac{2\pi}{\hbar^{f+1}} \int \ud^f J    
 \left. \frac{\partial H}{\partial I} \right|_{I=I(E,\vJ)}  ^{-1} 
     \frac{1}{4 \cosh^2 (\pi I(E,\vJ)/\hbar)}      \,.
\end{equation}
This term can easily be interpreted from
considering its integrated version 
\begin{equation} \label{eq:def_N_0}
N_0(E) = \int^E_{-\infty} \ud E' \, n_0(E')
= \frac{1}{\hbar^f} \int \ud^f J    
     \frac{1}{1+\ue^{-2\pi I(E,\vJ)/\hbar}}     \,. 
\end{equation}
In the semiclassical limit, $ \hbar \to 0$, the integrand can be
viewed as a characteristic function on the action space region
$I(E,\vJ)>0$.  The integral in \eqref{eq:def_N_0} hence gives the
action space volume enclosed by the surface $I(E,\vJ)=0$, and
accordingly $N_0(E)$ is given by the classical flux divided by the
elementary volume $(2\pi \hbar)^f$, which agrees with the mean
number of states of the activated complex to energy $E$
\cite{SchubertWaalkensWiggins06} (see Fig.~\ref{fig:momentummap}).
The term $n_0(E)$ is the corresponding differential version, i.e. the
mean density of states of the activated complex at energy $E$.


\emph{The oscillatory term.---}
To compute the terms $n_{\bf m}(E)$ for ${\bf m}\ne 0$ we use \cite{PrudnikovBrychkovMarichev86}
\begin{equation}
\frac{1}{4\cosh^2(\pi x)}=\frac{1}{(2\pi)^2}\int_{-\infty}^{\infty}\ud y \frac{y/2}{\sinh(y/2)}\ue^{-\ui yx}
\end{equation}
to rewrite \eqref{eq:def_n_m} as
\begin{equation}\label{eq:nm-ext}
\begin{split}
n_{\vm}(E) = \frac{\ue^{-\ui\pi \vm\valpha/2}}{2\pi\hbar^{f+1}}&\int\ud y\int_{J\geq 0} \ud^f J\bigg(\frac{\pa H}{\pa I}\bigg)^{-1}_{I(E,\vJ)} \times \\
& \frac{y/2}{\sinh(y/2)}\ue^{\ui[2\pi \vm\vJ-yI(\vJ,E)]/\hbar}\,.
\end{split}
\end{equation}
This  integral can be evaluated by the method of stationary phase. 
The stationary phase conditions are 
\begin{equation}\label{eq:stat-phase-cond}
2\pi \vm=y\nabla_{J}I_0(E,\vJ)\,\, ,\qquad  I_0(E,\vJ)=0\,,
\end{equation}
and by differentiating $H_0(I_0(E, \vJ),\vJ)=E$ we obtain
\begin{equation}
2\pi \vm=-\frac{y}{\lambda(\vJ)} \vomega(\vJ)\, \, ,\quad \text{where}\quad \lambda(\vJ):=\frac{\pa H_0}{\pa I}(0,\vJ)\, .
\label{eq:15}
\end{equation}
The second condition in \eqref{eq:stat-phase-cond} restricts $\vJ$ to
the energy surface of the activated complex $\Sigma_E$ defined in
\eqref{eq:def_Sigma_E}. The first conditions then fixes a point
$\vJ_{\vm}$ on $\Sigma_E$ (or a finite number of points $\vJ_{\vm,i}$)
by requiring that the frequency vector $\vomega(\vJ)$ at $\vJ_{\vm}$
is proportional to $\vm$.  By \eqref{eq:orbits} this means that the
torus corresponding to $\vJ_{\vm}$ is \emph{resonant}, and by
\eqref{eq:15} we have
\begin{equation}
\abs{y}=2\pi\frac{\lambda}{\abs{\vomega}}   \abs{\vm}\, ,
\end{equation}
where  $y<0$ ($y>0$) if $\vm$ and $\vomega$ are parallel (anti-parallel). 
Here all functions of $\vJ$ are evaluated at $\vJ_{\vm}$. 
Let $Q$ be the $(f+1)\times (f+1)$ matrix of second derivatives of the 
phase function in \eqref{eq:nm-ext} evaluated at $\vJ_{\vm}$ and $y$, and $\beta$ its signature. 
We then find for  $n_{\vm}(E)$,
\begin{equation}
 \frac{(2\pi)^{\frac{f-1}{2}}\ue^{-\ui[\pi \vm \valpha /2 + 2\pi \abs{\vm} \lambda I_1/\abs{\vomega} +\pi\beta/4]}}{\hbar^{\frac{f+1}{2}}\lambda\sqrt{\abs{\det Q}}}\frac{y/2}{\sinh(y/2)}\ue^{2\pi\ui  \vm\vJ_{\vm}/\hbar}\,.
\end{equation}

To evaluate  the determinant of $Q$  it is useful to introduce the curvature tensor $K$ of 
$\Sigma_E$. 
Let $\ve_1,\ve_2,\ldots ,\ve_{f-1}$ be $f-1$ orthogonal unit vectors 
which are tangent to $\Sigma_E$ at $\vJ$. Noting that $\Sigma_E$ is the hypersurface $I_0(E,\vJ)=0$ we can write  the components of $K$ at $\vJ_\vm$  
as 
\begin{equation}
K_{ij}=-\frac{1}{\abs{\nabla_\vJ I_0}}\ve_i\cdot I_0''\ve_j=-\frac{1}{\abs{\nabla_\vJ H_0}} \ve_i\cdot H_0''\ve_j\,,
\end{equation}
where $I_0^{''}$ and $H_0''$ denote the matrices of second derivatives with respect to $\vJ$. 
Let $\ve_1$ be the unit vector parallel to $\nabla_\vJ H_0=\vomega$.
Then in the basis of the $\ve_j$ the matrix $Q$ becomes 
\begin{equation}\label{eq:Q-sim}
Q=
\begin{pmatrix} 0 & \abs{\vomega}/\lambda & 0^T \\ \abs{\vomega}/\lambda &  -y\ve_1 I_0'' \ve_1 &\va^T \\
0 & \va & y\abs{\nabla_\vJ I_0}K 
\end{pmatrix}\,\, ,
\end{equation}
where $\va$ has components $\ve_1 I_0''\ve_j$.
The determinant of this matrix can be evaluated straightforwardly, but to determine as well 
the signature it is useful to rewrite it as follows. Let $A$ be the upper left $2\times 2$ block of 
\eqref{eq:Q-sim}, $D=y\abs{\nabla I_0}K$ 
and $B=\begin{pmatrix} 0 & \va\end{pmatrix}$, then if $\det K\neq 0$ we can form  
\begin{equation*}
\begin{pmatrix} A & B^T\\ B & D\end{pmatrix}
=\begin{pmatrix} I & B^T D^{-1}\\ 0 & I\end{pmatrix}\begin{pmatrix} A-B^TD^{-1}B & 0 \\ 0 & D\end{pmatrix}\begin{pmatrix} I & 0\\ D^{-1}B  & I\end{pmatrix}\,.
\end{equation*}
By the special structure of $B$ we find that $B^TD^{-1}B=\begin{pmatrix} 0 & 0\\ 0& c\end{pmatrix}$ for some number $c$. Hence  
 $\det(A-B^TD^{-1}B)=-\abs{\vomega}^2/\lambda^2<0$ and so  $A-B^TD^{-1}B$ has signature 
 $0$. The signature $\beta$ of $Q$ is thus determined by $D=y\abs{\nabla_\vJ I_0}K$ and we find 
 \begin{equation}
\beta=\sign y\sign K\,\, ,
\end{equation}
and with $y\nabla_\vJ I_0=2\pi \vm$, by \eqref{eq:stat-phase-cond},  the determinant is 
\begin{equation*}
\sqrt{\abs{\det Q}}=(2\pi \abs{\vm})^{\frac{f-1}{2}}\sqrt{\abs{\det K}}\, \abs{\vomega}/\lambda 
\end{equation*}
evaluated at $\vJ=\vJ_{\vm}$.  

We notice that if $\vJ_{\vm}$ and $y$ are a solution to the 
stationary phase condition for $\vm$, then $\vJ_{\vm}$ and $qy$ are a solution for $q\vm$ for any 
$q\in \Z\backslash \{0\}$. 
It is natural to choose  $\vmu\sim \vm$ with positive coprime components 
and   combine the two terms with $q\vmu$ and $-q\vmu$. This way the $n(E)$ contribution of 
the $q\text{th}$ repetition of a resonant torus with $\vomega\sim\vmu$ is given by
\begin{equation} \label{eq:def_n_mu_q}
\begin{split}
&n_{\vmu,q} (E) = \frac{2\pi}{\hbar^{(f+1)/2}} \frac{\lambda}{\sinh\big( \pi q \frac{|\vmu|}{|\vomega|} \lambda\big)} \times \\
&\frac{\cos \big( q  (2\pi    \vmu  \cdot \vJ/\hbar     - \pi  \vmu \valpha  /2  - 2\pi \abs{\vmu} \lambda I_1/\abs{\vomega}  ) + \pi  \beta/4  \big) }{ q^{(f-3)/2}  |\vmu |^{(f-3)/2}   |\vomega|^2 \sqrt{|\det K(\vJ^\vmu)|}}\,.
\end{split}
\end{equation}

\begin{figure}
\begin{center}
\includegraphics[angle=0,width=7cm]{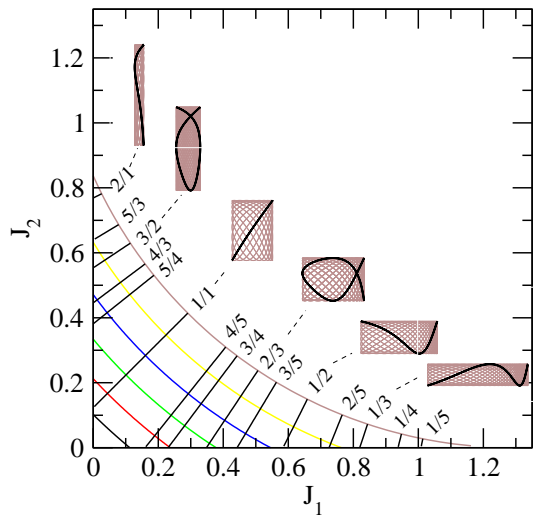}
\end{center}
\caption{\label{fig:esurf_xfig}
Energy surfaces with resonance lines $\mu_1/\mu_2$
of the activated complex which consists of two Morse oscillators.
The insets show  the resonant tori with $\mu_1,\mu_2\le 3$ projected to the configuration space of the oscillators.
}
\end{figure}
\begin{figure}
\begin{center}
\includegraphics[angle=0,width=6.5cm]{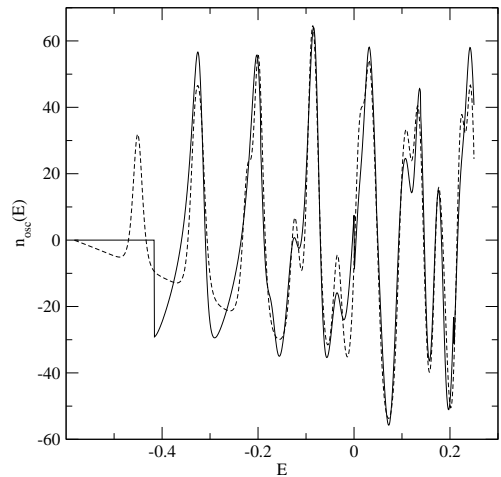}
\end{center}
\caption{\label{fig:dcum_dE}
Exact (dashed line) and periodic orbit approximation (solid line) of the energy derivative of the cumulative reaction probability including resonant tori with $\mu_1,\mu_2\le 3$.
}
\end{figure}

\emph{Example.---}
We consider the example of a system composed of an Eckart barrier and two Morse oscillators. 
Its quantum normal form Hamiltonian is given by \cite{SchubertWaalkensWiggins06}
\begin{equation}
\hat{H} = ( \frac{\pi}{a_0}\sqrt{\frac{V_0}{2m}}+\hat{I})^2  - \sum_{k=1}^2(\sqrt{D_k} - \frac{a_k}{\sqrt{2m}} \hat{J}_k )^2   \,.
\end{equation}
Here $H=H_0$, and hence $I_1=0$. The frequencies are
\begin{equation}
\Omega_k= \partial_{J_k} H_0 = a_k \sqrt{\frac{2D_k}{m}}-\frac{a_k^2}{m}J_k\, , \quad k=1,2\,,
\end{equation}
and 
\begin{equation}
\lambda=\left. \frac{\partial H_0}{\partial I}  \right|_{I=0}=  \frac{2\pi}{a_0} \sqrt{\frac{V_0}{2m}}  \,. 
\end{equation}
We choose $D_2=a_1=a_2=1$,  $D_1=5/6$, $a_0=4\pi$, $V_0=5/4$, and $\hbar=0.1$.
Figure~\ref{fig:esurf_xfig} shows energy surfaces $\Sigma_E$ of the activated complex consisting of the two Morse oscillators together with some resonance lines $\Omega_1/\Omega_2=\mu_1/\mu_2$. The sign of the curvature matrix is $\beta=-1$. The exact cumulative reaction probability, and its derivative $n(E)$ can be computed analytically for this system \cite{SchubertWaalkensWiggins06}. Its oscillatory part,
$n_{\text{osc}}=n(E)-n_0(E)$,  is  shown together with its approximation by the periodic orbit sum over the terms 
\eqref{eq:def_n_mu_q}  for $\mu_1,\mu_2\le 3$ in Fig.~\ref{fig:dcum_dE}. 

\emph{Conclusions.---}
In this letter we derived a periodic orbit formula for the cumulative
reaction probability, and demonstrated its applicability for a simple
example.  In the limit $\lambda\to0$ (no tunneling through the
potential barrier) our periodic orbit formula reduces to the
Berry-Tabor trace formula for the density of states of the activated
complex. In the general case $\lambda\ne0$, our periodic orbit formula
is (as opposed to the Berry-Tabor trace formula) {\it absolutely
  convergent} due to an additional factor which leads to an
exponential damping of contributions of long periodic orbits.
Although we incorporated only six periodic obits (and their
repetitions) in our example the agreement with the exact result is
already very good. This is even more impressive as we have so far only
taken into account simple stationary points associated with resonant
tori, and no isolated and ghost orbits which would naturally arise in
a more elaborate uniform approximation
\cite{BerryTabor76,Richens82}. Similarly, the integral associated with
the reaction direction can be cast into a periodic orbit sum over the
instanton orbits \cite{Miller77} extending the applicability of our
periodic orbit formula to energies below the saddle energy. These
aspects will be discussed in more detail in a longer version of this
letter.


\emph{Acknowledgments.---}
This work was supported by EPSRC under Grant No. EP/E024629/1 and ONR under Grant No. N00014-01-1-0769.





\end{document}